\documentclass[twocolumn,prm,showpacs,floatfix,superscriptaddress]{revtex4-2}

\usepackage{placeins}
\usepackage{graphicx}
\usepackage{dcolumn}
\usepackage{bm}
\usepackage{amssymb}
\usepackage{amsmath}
\usepackage{amsfonts}
\usepackage{subfigure}
\usepackage{hyperref}
\usepackage{xcolor}
\usepackage{siunitx}
\usepackage{array}

\providecommand{\U}[1]{\protect\rule{.1in}{.1in}}
\newcommand{\pib}{PrIr$_3$B$_2$}
\usepackage{xr-hyper}

\begin{document}
\title{Tunability of the structural and magnetic transition in kagome material: PrIr$_3$B$_2$ }
\author{Gayathri V}
\email{gayathriv@iitpkd.ac.in}
\affiliation{Department of Physics, Indian Institute of Technology Palakkad, Palakkad-678623, Kerala, India.}

\author{Irshad K A}
\affiliation{Elettra-Sincrotrone Trieste S.C.p.A., S.S. 14, Km 163.5 in Area Science Park, Basovizza 34149, Italy.}

\author{Sathishkumar M}
\affiliation{Department of Physics, Indian Institute of Technology Palakkad, Palakkad-678623, Kerala, India.}

\author{Boby Joseph}
\affiliation{Elettra-Sincrotrone Trieste S.C.p.A., S.S. 14, Km 163.5 in Area Science Park, Basovizza 34149, Italy.}

\author{S. Manni}
\email{smanni@iitpkd.ac.in}
\affiliation{Department of Physics, Indian Institute of Technology Palakkad, Palakkad-678623, Kerala, India.}
\date{\today}

\begin{abstract}
We report the temperature and pressure tunability of an unusual structural transformation associated with a two-step metal-insulator-metal (MIM) transition in the kagome lattice compound PrIr$_3$B$_2$ using synchrotron X-ray powder diffraction. At ambient conditions of temperature and pressure, the monoclinic ($C2/m$) and the hexagonal ($P6/mmm$) phases coexist as twinned structure in the crystal. As the temperature (pressure) is decreased (increased), PrIr$_3$B$_2$ converts fully to monoclinic structure at $T =$ 280~K (at ambient pressure) and $P =$ 1.2~GPa (at room temperature). Temperature dependence of the  monoclinic structure at ambient pressure presents complex evolution of lattice parameters with weak but clearly discernable anomalies at \SI{\sim 250} {\K} and \SI{\sim 110} {\K}, which are correlated with the second MIM transition and the linear to nonlinear temperature-dependent resistivity crossover, respectively. These anomalies are likely due to some charge order state causing a partially gapped Fermi surface. The magnetic phase diagram of PrIr$_3$B$_2$ is also investigated from anisotropic measurements. At 10~K, a superzone gap opens near the antiferromagnetic transition, which does not close even in the polarized state. From the tunability of the crystal structure and magnetic and electronic ground state, promising electronic orders are indicated in this kagome metallic magnet.

\end{abstract}

\maketitle

\hspace{5.2in}

\section{Introduction}
The ternary compound series, RT$_3$X$_2$ (R = rare earth element, T = $4d$ or $5d$ transition metal, X = Si, B), were discovered in 1980 \cite{Ku1980} and exhibit diverse exotic phenomena like magnetism, superconductivity, structural transition, charge-density wave, etc., based on the nature of the rare earth elements involved \cite{Dhar1981, Malik1983, Langen1987, Berger2001, Okubo2003, Kubota2013, Manni2019, Gui2022, Chakrabortty2023, Chaudhary2023, Xiao2025, Kral2026}. In these compounds, the transition metal atoms form a kagome lattice. Kagome lattices provide an interesting playground for investigating the interplay among various exotic phenomena arising from their non-trivial band topology, geometric frustration, and strong electronic correlations \cite{Ye2018, Ortiz2019, Yin2022, Negi2025}. Nevertheless, the physical properties exhibited by RT$_3$X$_2$ systems were not addressed in the context of their kagome lattices until recently. LaRu$_3$Si$_2$, YRu$_3$Si$_2$, LaRh$_3$B$_2$, and YRu$_3$B$_2$ are a few members reported so far, in which the observed unconventional properties have been attributed to electron correlations from the flat bands associated with their kagome lattices \cite{Gong_2022, Chakrabortty2023, Chaudhary2023, Gaggl2026}. 

Among the RT$_3$X$_2$ series, CeRh$_3$B$_2$ is an exceptional ferromagnet with a Curie temperature ($T_{\rm C}$) of 120~K, being the highest reported $T_{\rm C}$ so far for any Ce-based compound devoid of magnetic elements \cite{Dhar1981}. This anomalous ferromagnetism observed in CeRh$_3$B$_2$ is reported to originate from the coexistence of both localized and delocalized band-like $4f$-states arising from the hexagonal symmetry, resulting in the ordering of Ce local moments, coupled together by unusually strong Ce $4f$-conduction electron hybridisation \cite{Shaheen1985, Shaheen1986}. High-pressure studies on CeRh$_3$B$_2$ revealed an initial increase in $T_{\rm C}$, passing through a maximum and finally rapidly falling at higher pressures \cite{Cornelius1994}. LaRh$_3$B$_2$ is so far the only compound among the RRh$_3$B$_2$ series reported to exhibit superconductivity (at \SI{\sim 2.6} {\K}) \cite{Misemer1984}. The Os-analogue, LuOs$_3$B$_2$, has been reported to exhibit superconductivity at \SI{\sim 4} {\K} \cite{Ku1980}. Ru-analogue of the family, YRu$_3$B$_2$, has recently been reported to exhibit bulk superconductivity at 0.7~K \cite{Gaggl2026}. Among the Ir-analogues, CeIr$_3$B$_2$, undergoes a structural transition from the room-temperature monoclinic to a hexagonal structure at ~395~K and also possesses a relatively high $T_{\rm C}$ of 41~K \cite{Kubota2013}. The superconductivity or unconventional magnetism exhibited by these systems is reported to arise from the non-trivial band topology associated with the kagome lattice of the non-magnetic transition metal. The Ir-analogue of the family is particularly interesting candidate to probe for hosting such strongly correlated electronic states due to the spin-orbit coupling associated with its higher nuclear charge compared to other transition metal atoms \cite{Uzunok2025}. Superconductivity has been discovered in some of the Ir-analogue like LaIr$_3$B$_2$ and ThIr$_3$B$_2$ \cite{Ku1980}. 

PrIr$_3$B$_2$ (PIB) is yet another member of the RT$_3$X$_2$ family having an Ir kagome lattice. PIB is reported to crystallize in the hexagonal CeCo$_3$B$_2$-type ($P6/mmm$) structure at room temperature \cite{SOLOGUB2003, Manni2019, Sharma2024}. The studies on the single crystal of PIB reported a first-order transition at 280~K with a second-order antiferromagnetic transition at 10~K \cite{Manni2019}. Neutron diffraction measurements performed on the polycrystals of PIB confirmed the long-range ordering of Pr$^{3+}$ spins, with no apparent magnetic moment on the Ir sites \cite{Sharma2024}. More interestingly, the electrical resistivity measurement performed on the single crystal of PIB exhibited an anomalous two-step metal-insulator-metal (MIM) transition in the temperature range 250 to 280~K, unlike the other members of the RT$_3$X$_2$ series \cite{Manni2019}. The origin of this two-step MIM transition is not yet understood. The temperature-dependent powder diffraction measurements performed on polycrystalline PIB in the previous reports using a lab source were inconclusive in giving a clear understanding of their structural evolution \cite{Sharma2024}. In this work, we have investigated the evolution of structural properties of PIB under extreme conditions of temperature and pressure using synchrotron radiation to understand the origin of this two-step MIM transition. Multiple structural changes were observed, correlating with anomalous electrical transport in this material. We have also mapped the magnetic phase diagram of PIB from magnetic field-dependent magnetization, electrical and thermodynamic measurements on an oriented single crystal.

\section{Experimental Methods}
The single crystal of PrIr$_3$B$_2$ (PIB) was synthesised by the Czochralski pulling method in a tetra arc furnace, as described in Ref. \cite{Manni2019}. Differential scanning calorimetry (DSC) measurements were performed on a crystal weighing \SI{\sim 11.8}{\milli\gram} using a NETZSCH DSC 204 F1 PHOENIX, equipped with a refrigerated cooling system under an inert atmosphere of argon. The temperature-dependent DSC heat flow (both heating and cooling cycles) and DSC specific heat capacity (heating cycle) were measured between the temperature ranges 200~K and 450~K, at a scanning rate of 10~K/min. We have used sapphire as a standard reference for this DSC measurement. Resistivity measurements were performed using the four-probe method in the current-reversal mode using an Oxford Teslatron cryostat and a Quantum Design Physical Property Measurement System (PPMS) in the temperature range of 4 to 300~K. Temperature and field dependent magnetization and heat capacity measurements were done using a Quantum Design Magnetic Property Measurement System (MPMS 3) SQUID magnetometer setup and PPMS setup, respectively.  Temperature and pressure-dependent synchrotron X-ray diffraction (XRD) measurements were performed at the Xpress beamline of Elettra Sincrotrone, Trieste, Italy. Crushed fine powders of the single crystal were used for the XRD measurements. An X-ray of wavelength \SI{0.4953}{\angstrom} was used as the source in the transmission Debye-Scherrer mode. Dectris PILATUS3 S 6M detector was used to collect the diffracted beam. The obtained two-dimensional diffraction images were converted to one-dimensional $2\theta$ vs intensity patterns using the Dio
ptas GUI software \cite{Prescher2015}. The temperature-dependent XRD measurements were carried out inside a Lakeshore-Janis closed-cycle helium cryostat in the temperature range 310 to 50~K. Lead (Pb) was used as an internal temperature calibrant. The high-pressure XRD measurements were carried out using the membrane diamond anvil cell, and the applied pressure was varied from 0 to 10~GPa. Methanol-ethanol mixture in the ratio 4:1 was used as the pressure-transmitting medium and a tiny ruby ball embedded inside the sample chamber was used for the accurate pressure determination. To understand the reversibility of the transition, additional data points were collected in the reverse cycle. Rietveld refinement was carried out to extract the structural information using General Structure Analysis Software (GSAS)-II \cite{Toby2013}. The equation of state fitting of the obtained pressure-volume ($P-V$) curve was carried out using EoSFit7-GUI software \cite{gonzalez2016eosfit7}.

\section{Results and Discussions}

\subsection{Differential Scanning Calorimetry}
\begin{figure}[htbp]
\centering
\includegraphics[width=\columnwidth]{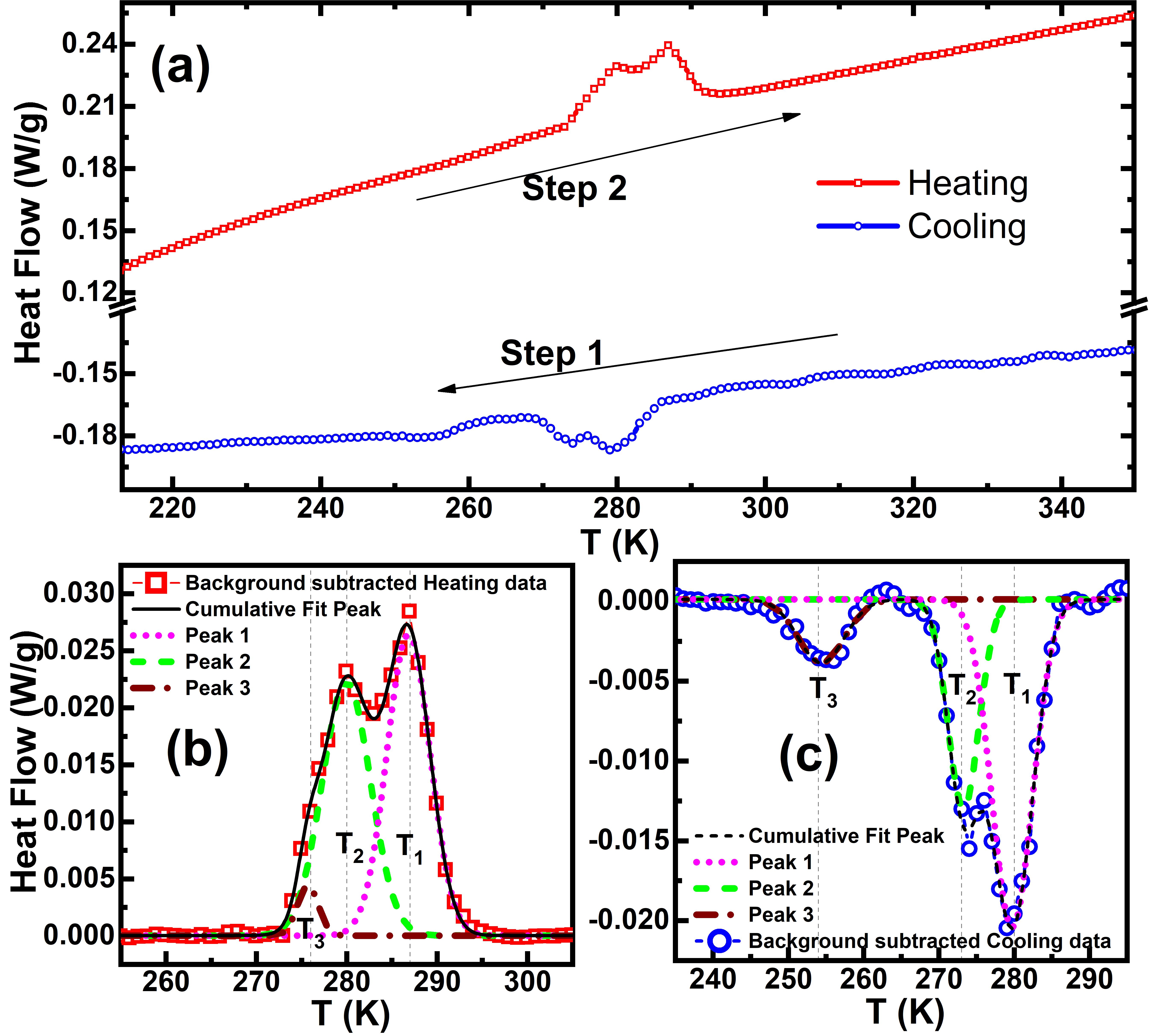}
\caption{(a) DSC heat flow in the heating and cooling cycle. The arrows indicate the sequence of measurements, where cooling data is collected in step 1 followed by the heating data in step 2. Presence of three peaks after the background subtraction and deconvolution of the (b) heating and (c) cooling cycles.}
\label{fig:DSC_2a_2b}
\end{figure}

The thermal analysis using DSC specific heat capacity measurement performed on the single crystal indicates a broad peak between 285~K and 295~K with a jump of around 200~J/K-mol, as shown in Fig.~\ref{fig:DSC_2a} of the supplementary material. Fig.~\ref{fig:DSC_2a_2b}~(a) illustrates the DSC heat flow curves in the cooling and heating cycles. Fig.~\ref{fig:DSC_2a_2b}~(b) and (c) highlights the DSC heating and cooling curves, respectively, after subtracting the baseline and deconvolution of the peaks. A careful look at the DSC curve in the cooling cycle reveals three distinct peaks 1, 2, and 3 at $T_1$, $T_2$, and $T_3$, respectively (see Fig.~\ref{fig:DSC_2a_2b}~(c)). The peaks at $T_1$ and $T_2$ are sharper and only around 5~K apart, while the peak at $T_3$ is much weaker and more than 25~K apart. A clear shift of the peaks is observed in the heating cycle, where the peak at $T_3$ now translates to a higher temperature and appears as a shoulder of the low-temperature side of the peak at $T_2$ (see Fig.~\ref{fig:DSC_2a_2b}~(b)). After subtracting the baselines and deconvoluting the peaks, the peak positions are found to be $T_1$=287~K, $T_2$=280~K, and $T_3$=276~K, in the heating cycle. Similar peak positions in the cooling cycles were observed at $T_1$=280~K, $T_2$=273~K, and $T_3$=254~K, respectively. 

\subsection{Electrical resistivity}

To understand the origin of the transitions, the electronic properties of the crystal were studied by performing temperature-dependent resistivity measurement on an unoriented crystal. Fig.~\ref{fig:RT_insets_ab_linear fit} of the supplementary material portrays the temperature-dependent electrical resistivity ($\rho(T)$ vs $T$) measurement. The value of resistivity at 4~K was found to be \SI{130}{\micro\ohm\centi\meter}, which is comparable to that of the other RT$_3$X$_2$ systems \cite{Kubota2013} and yielded a residual resistivity ratio ($\rho$(300 K)/$\rho$(4 K)) of 1.35. The electrical resistivity measurements reveal an anomalous two-step metal-insulator-metal (MIM) transition in the temperature range 250 to 280~K, as highlighted in inset (a) of Fig.~\ref{fig:RT_insets_ab_linear fit} of the supplementary material, similar to that reported earlier for current along the $[10\overline{1}0]$ direction \cite{Manni2019}. The dip in the $d\rho/dT$ vs $T$ curve was used to locate the transition temperatures. The two transitions in the heating cycle were identified to be at 280~K and 273~K. In the cooling cycle, they were found to be located at 279~K and 258~K. The second transition observed at \SI{\sim 273}{\K} in the heating cycle is highly hysteretic and much broader in comparison to the first transition observed at \SI{\sim 280}{\K}. These transitions seen in the transport measurement can be related to those detected in the DSC measurement. The first sharp transition at \SI{\sim 280}{\K} in the resistivity may originate from the merging of the closely spaced $T_1$ and $T_2$ transitions observed in DSC, whereas the second broad and hysteretic feature may correspond to the $T_3$ transition seen in DSC. In the following text, the first and the second transitions seen in transport measurement will be addressed as $T_{1,2}$ and $T_3$, respectively. Below the two-step MIM transition, the crystal exhibits a metallic behaviour. Interestingly, below $T_3$, the $\rho (T)$ exhibits a linear temperature dependence down to \SI{\sim 110}{\K}, i.e., at temperatures well below the Debye temperature ($\theta_D$) of the system. This suggests the possibility of anomalous electron-phonon scattering mechanisms originating from strong electron-electron correlation or partially gapped Fermi surface below $T_3$. However, further detailed investigation is required to understand this. On further cooling, it shows a deviation from this linear behaviour below 110~K, as shown in inset (b) of Fig.~\ref{fig:RT_insets_ab_linear fit} of the supplementary material. Additionally, a resistive upturn is observed at \SI{\sim 10}{\K}, associated with the magnetic transition exhibited by the crystal (discussed in a later section) \cite{Manni2019}.

\subsection{Temperature and pressure tuning of crystal structure}
To delineate the structural origin of these transitions, temperature and pressure-dependent XRD measurements were performed using a synchrotron radiation. 

\subsubsection{Structural properties at ambient conditions (\SI{\sim 300}{\K} and \SI{\sim 0}{\GPa})}

\begin{figure*}[htbp]
\centering
\includegraphics[width=\textwidth]{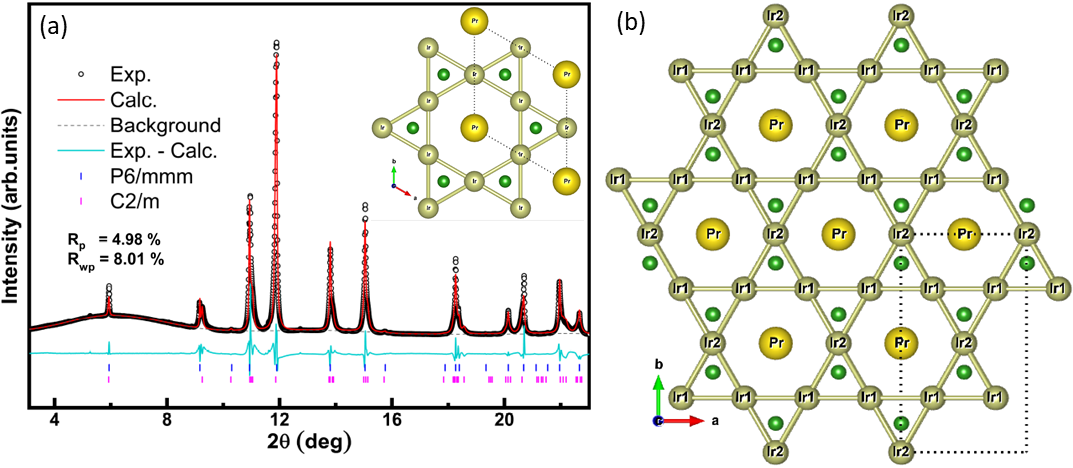}
\caption{(a) Results of the Rietveld refinement analysis of the synchrotron powder XRD pattern of the powdered PIB single crystal at ambient pressure and room temperature. Inset here shows the crystallographic structure of the hexagonal phase along the $ab$ plane revealing Ir atoms forming the kagome lattice. Green spheres are the B atoms. Dotted rectangle indicates the unit-cell projection. (b) The crystallographic structure along the $ab$ plane of the monoclinic phase.}
\label{fig:300K_ambient}
\end{figure*}

Fig.~\ref{fig:300K_ambient}~(a) presents the synchrotron XRD pattern obtained for the powdered single crystal of PIB at ambient conditions (\SI{\sim 300}{\K} and \SI{\sim 0}{\GPa}) along with the Rietveld refinement results. Based on the previous reports on similar RT$_3$X$_2$ systems \cite{Manni2019, Kubota2013}, PIB is expected to exist in the hexagonal structure (with space group $P6/mmm$) at high temperature. However, a single-phase refinement using hexagonal $P6/mmm$ phase alone did not yield a good fit. It was observed that the peaks in the ambient XRD pattern were asymmetric with shoulders that cannot be accounted by the Bragg peaks of the hexagonal structure alone, suggesting the coexistence of at least another phase. This can be clearly visualized by considering the peak at $2\theta$ = 9.14$^{\circ}$. At this $2\theta$ position, the hexagonal phase has only one Bragg peak corresponding to $(001)$ reflection. Nevertheless, an additional well-resolved Bragg peak has been observed adjacent to the hexagonal $(001)$ reflection. To account for the additional reflection, the inclusion of a secondary phase was necessary. The primary consideration was monoclinic ($C2/m$) structure, consistent with previous literatures \cite{Manni2019, Sharma2024}. Hence, a quantitative two-phase Rietveld refinement was performed to extract the structural properties of PIB at ambient conditions. As can be appreciated from Fig.~\ref{fig:300K_ambient}~(a), such a two-phase Rietveld refinement yielded a reasonably good agreement of the XRD pattern at \SI{\sim 300}{\K} and \SI{\sim 0}{\GPa}. We note that the diffraction data had high preferential orientation, and a minimal intensity correction using a spherical harmonics model is employed to address it. Thus, our synchrotron XRD data clearly reveals that at ambient conditions, the hexagonal ($P6/mmm$) and monoclinic ($C2/m$) phases coexist in PIB. From the analyses, we obtained the phase fractions of the monoclinic and the hexagonal phases to be \SI{\sim 48}{\percent} and \SI{\sim 52}{\percent}, respectively. The structural model of the PIB highlighting the kagome lattice arrangement of the Ir atoms are also shown in Fig.~\ref{fig:300K_ambient}. As can be readily seen from the inset in Fig.~\ref{fig:300K_ambient}~(a), in the case of hexagonal structure, the Ir atoms arrange well in the kagome lattice. In the monoclinic structure, Ir atoms take two distinct crystallographic positions, Ir1 and Ir2 (see Fig.~\ref{fig:300K_ambient}~(b) and Table~\ref{tab:PrIr3B2}). However, it still keeps the kagome arrangement due to the similar bond distances for the Ir1-Ir1 (\SI{2.769}{\angstrom}) and Ir1-Ir2 (\SI{2.772}{\angstrom}). Compared to the hexagonal phase, B atoms (green spheres) in the monoclinic phase are seen to be preferentially close to one of the Ir sites. The crystallographic data for the two phases from our refinement analyses are provided in Table~\ref{tab:PrIr3B2} of the supplementary material. 

\subsubsection{Structural properties at low temperature}

\begin{figure*}[htbp]
\centering
\includegraphics[width=\textwidth]{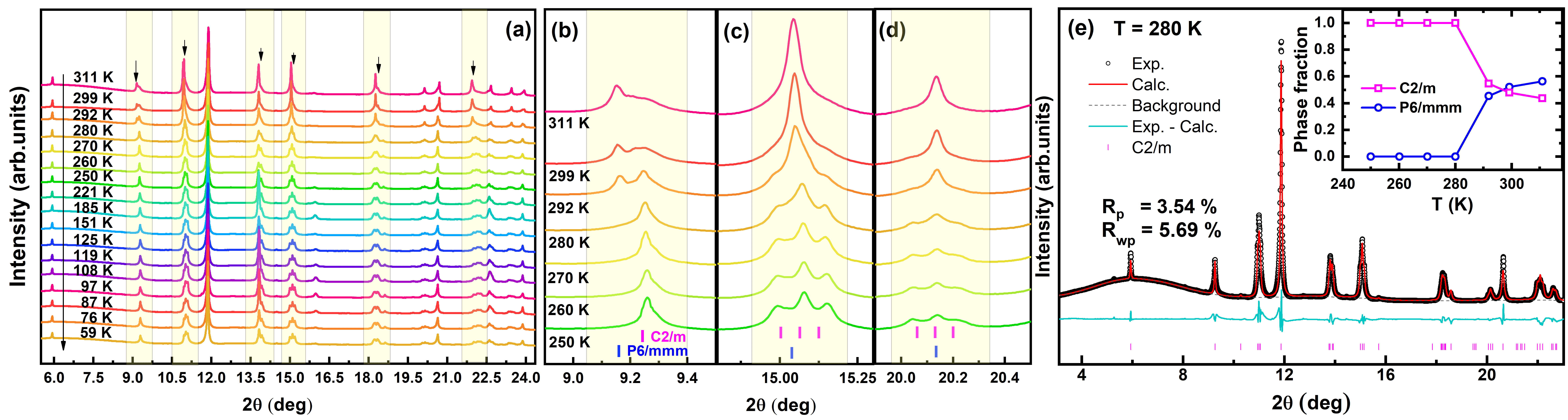}
\caption{(a) Evolution of the XRD pattern with temperature. Structural transition from mixed phase to purely monoclinic phase highlighted around the peak at (b) 9.2$^{\circ}$, (c) 15$^{\circ}$, (d) 20.15$^{\circ}$ near the transition temperature. (e) Synchrotron powder XRD pattern obtained for powdered single crystal at 280 K at ambient pressure. The inset in(e) shows the variation of phase fractions of two coexisting phases with temperature in the cooling cycle in PIB.}
\label{fig:PIB_LTrun_PeakEvolution}
\end{figure*}

\begin{figure*}[ht!]
 \includegraphics[width=\textwidth]{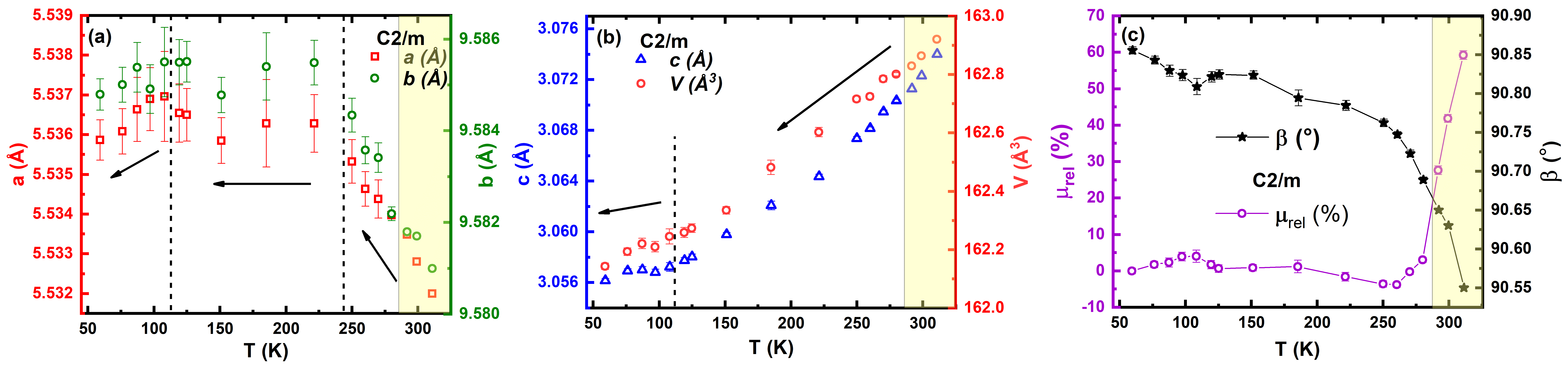}
\caption{Variation of the lattice parameters of monoclinic ((a)-(c)) phase as a function of temperature obtained from the Rietveld refinement of the temperature-dependent XRD pattern. The shaded region in (a), (b), and (c) indicate the variation of the lattice parameters of monoclinic structure in the mixed phase region.}
\label{fig:S1_Refined_parameters_Tdep}   
\end{figure*}

To determine the structural evolution, temperature-dependent XRD measurements were performed at ambient pressure. Temperature-dependent XRD patterns in the temperature range 310 to 50~K are shown in Fig.~\ref{fig:PIB_LTrun_PeakEvolution}~(a). By observing the peaks at $2\theta$ = 9.2$^{\circ}$, 11$^{\circ}$, 14$^{\circ}$, 15$^{\circ}$, 18.1$^{\circ}$, 20.15$^{\circ}$, and 22$^{\circ}$, a gradual and progressive structural transition from the high-symmetry hexagonal ($P6/mmm$) to the low-symmetry monoclinic ($C2/m$) phase can be visualized by the disappearance of the peaks corresponding to the high-symmetric structure and emergence of the split peaks corresponding to the lower-symmetric structure, as highlighted in Figs.~\ref{fig:PIB_LTrun_PeakEvolution}~(b)-(d). For instance, the peak at 20.15$^{\circ}$ corresponding to the $(301)$ reflection of the $P6/mmm$ structure gradually splits into the $(331)$, $(-331)$, and $(061)$ peaks of the $C2/m$ structure on cooling (see Fig.~\ref{fig:PIB_LTrun_PeakEvolution}~(d)). The peak at $2\theta$ = 9.14$^{\circ}$ corresponding to the $(001)$ reflection of the $P6/mmm$ structure disappears on cooling (see Fig.~\ref{fig:PIB_LTrun_PeakEvolution}~(b)). In addition to this peak splitting and peak disappearance, the temperature evolution of the Rietveld refinement parameters like phase fraction, microstrain, lattice parameters, etc. were also inspected to identify the transition temperatures. The temperature evolution of the phase fraction of the two phases extracted from refinement of the XRD patterns in the cooling cycle is shown in the inset of Fig.~\ref{fig:PIB_LTrun_PeakEvolution}~(e). With decreasing temperature from room temperature, there is a gradual increase in the phase fraction of the monoclinic phase in the crystal.  For  $T\leq$ 280~K, the crystal is in the single phase (monoclinic) region. The region $T >$ 280~K marks the mixed phase region, where the monoclinic and hexagonal phases coexist in a twinned structure. The representative XRD pattern at 310~K, refined with the two phases (monoclinic and hexagonal) and at 280~K, refined with a purely monoclinic phase, is shown in Fig.~\ref{fig:310K_refinement_PF_Heating} of the supplementary material and Fig.~\ref{fig:PIB_LTrun_PeakEvolution}~(e), respectively. The temperature evolution of various structural parameters of the monoclinic and the hexagonal phase is illustrated in Figs.~\ref{fig:S1_Refined_parameters_Tdep}~(a)-(c) and Fig.~\ref{fig:lattice_parameter_LT_Hex} of the supplementary material, respectively. To extract the structural parameters, two-phase refinement using monoclinic and hexagonal phases was employed for $T >$ 280~K and single-phase refinement using the monoclinic structure was used for $T\leq$ 280~K. The structural transition temperature can also be identified from the temperature evolution of the microstrain, as shown in terms of relative microstrain ($\mu_{\mathrm{rel}}$ (\si{\percent})) (calculated relative to the microstrain at 50~K) in Fig.~\ref{fig:S1_Refined_parameters_Tdep}~(c). The value of $\mu_{\mathrm{rel}}$ (\si{\percent}) is calculated using the relation \eqref{eq:Relative microstrain}, where $\mu_{\mathrm{T,P}}$ is the microstrain at a particular temperature or pressure and $\mu_{\mathrm{R}}$ is the microstrain at the reference temperature or pressure. The larger value of $\mu_{\mathrm{rel}}$ (\si{\percent}) at ambient obtained for both the phases from the refinement indicates that the inherent strain in the system stabilizes the twinned structure in the mixed phase region. As the temperature decreases, the $\mu_{\mathrm{rel}}$ (\si{\percent}) reduces until the structural transition into the purely monoclinic phase attains completion, as shown in Fig.~\ref{fig:S1_Refined_parameters_Tdep}~(c). This means that the transition into the stable monoclinic phase is accompanied by the strain relaxation. Thus, by correlating the disappearance of the peaks corresponding to the hexagonal phase at 280~K with the estimated phase fraction and the microstrain reduction, the structural transition into the monoclinic phase is found to be complete at 280~K, and a wide mixed phase region is identified where the two phases exist in a twinned structure at $T >$ 280~K. Furthermore, the temperature evolution of the angle $\beta$ suggests that across the structural transition, the monoclinic structure evolves by the shear distortion of the hexagonal structure along the $ab$-plane, as seen from the enhancement in the value of angle $\beta$ with decreasing temperature (see Fig.~\ref{fig:S1_Refined_parameters_Tdep}~(c)). The temperature evolution of the lattice parameters are illustrated in Figs.~\ref{fig:S1_Refined_parameters_Tdep}~(a) and (b) for the monoclinic structure, and Fig.~\ref{fig:lattice_parameter_LT_Hex} of the Supplementary material for the hexagonal structure.
With decreasing temperature, an overall decrease in the unit cell volume is observed due to the contraction of the lattice. Interestingly, while the $c$ lattice parameter of both the monoclinic and hexagonal phases decrease with temperature (positive thermal expansion (PTE)) in the entire temperature range of study, the $a$ and $b$ lattice parameters show steady increase with decreasing temperature (negative thermal expansion (NTE)) in the temperature range 250~K $<T\leq$ 310~K. This NTE observed along the $a$ and $b$ axes of the monoclinic structure can be correlated to the NTE along the $a$ axis of the hexagonal structure (see Figs.~\ref{fig:S1_Refined_parameters_Tdep}~(a) and (b) for the monoclinic structure, and Fig.~\ref{fig:lattice_parameter_LT_Hex} of the Supplementary material for hexagonal structure). Given the fact that, the value of $a$ and $b$ lattice parameters of the monoclinic structure is very close to that of the $a$ lattice parameter of the hexagonal structure, the NTE property could be inherited or retained across the phase transition due to the close structural relationship between the two phases (see Fig.~\ref{fig:300K_ambient}~(a) inset and Fig.~\ref{fig:300K_ambient}~(b)).

\begin{equation}
\mu_{\mathrm{rel}} = \frac{{\mu_{\mathrm{T,P}}-\mu_{\mathrm{R}}}}{\mu_{\mathrm{R}}} \times 100
\label{eq:Relative microstrain}
\end{equation}

Hence, at 280~K, PIB transforms to a completely monoclinic structure upon cooling from a mixed hexagonal and monoclinic structure. It is noteworthy that the transition from monoclinic to mixed-phase also happens at 280~K during the heating cycle (in 10~K step measurements, see Fig.~\ref{fig:PIB_LTrun_PeakEvolution_FWHM_1} and the inset of Fig.~\ref{fig:310K_refinement_PF_Heating} of the supplementary material). DSC measurement has shown maximum heat flow at around this temperature and a maximum change in resistivity is also observed there. Hence, the phase transition at $T_1$ = 280~K is clearly related to the structural transformation and very weakly hysteretic. This is very unusual for a first-order structural transition, which can only be justified by the broad transition due to inherent strain in the crystal. No noticeable structural anomalies/changes were detected around $T_2$, which is, anyway, very close to $T_1$.

At $T_3$, the crystal undergoes the broad hysteretic transition at \SI{\sim 250}{\K} in the cooling cycle, and at \SI{\sim 270}{\K} in the heating cycle as seen in the temperature-dependent resistivity and DSC measurements. Our structural analyses reveal that the PIB is in a purely monoclinic structure at this temperature. By carefully looking at the evolution of the $a$ and $b$ lattice parameters of the monoclinic structure below 280~K, as shown in Fig.~\ref{fig:S1_Refined_parameters_Tdep}~(a), it is observed that the $a$ and $b$ lattice parameters steadily increase with decreasing temperature up to \SI{\sim 250}{K}, indicating an NTE behaviour in the $ab$-plane for 250~K $\leq T \leq$ 310~K. Thereafter, the changes in the $a$ and $b$ lattice parameters remain almost negligible in the temperature range 110~K $< T <$ 250~K, showing a clear structural anomaly at \SI{\sim 250}{K}. Additionally, a second anomaly in the evolution of all the lattice parameters is seen at \SI{\sim 110} {\K}, as shown in Figs.~\ref{fig:S1_Refined_parameters_Tdep}~(a)-(c). This anomaly is more pronounced in the variation of the lattice parameters $a$, $b$, and $\beta$ of the monoclinic structure. Unlike their trend up to 110~K, the $a$ and $b$ lattice parameters start decreasing with further cooling below 110~K, indicating a PTE behaviour below 110~K. The values of thermal expansion coefficient ($\alpha_i$), as estimated using the relation \eqref{eq:Thermal_expansion} in different temperature ranges, are tabulated in Table~\ref{tab:structure} for comparison. Here, $i$=$a,b,c$-lattice parameter at temperature $T$, $i_0$ is the lattice parameter corresponding to the lowest temperature in a specific temperature range, and $\Delta T$ is the width of the temperature range. The anomalous deviation from linear behaviour observed in the transport measurements below 110~K must be coupled with this subtle changes in the lattice parameters observed around 110~K. Also, no significant changes are observed in the magnetic properties down to 50~K \cite{Manni2019}.

Hence, the transition at $T_3$ and the anomaly at 110~K are electronic in origin, accompanied by subtle changes in the thermal expansion coefficients without any further structural transition. These anomalies could be manifestation of a strong electron-phonon coupling, which modifies the lattice parameters of the crystal \cite{Langen1987}. By correlating the observed structural changes with that of the second MIM transition in resistivity and weak anomaly in DSC, the origin of $T_3$ can be associated with the formation of a possible charge density wave (CDW) state. This further explains the linear resistivity behaviour observed in the $T$-range 110~K $<T\leq$ 250 K, which is at temperatures well below the $\theta_D$ of the system. Reconstruction of Fermi surface (FS) with the presence of both gapped and ungapped regions in the CDW state was previously reported in GdTe$_3$ \cite{Regmi2023}. The linear resistivity behaviour in the $T$-range 110~K $<T\leq$ 250 K could be due to the presence of a shrunken FS as a consequence of CDW, wherein now only tiny pockets of FS are available for conduction. No superlattice peaks were observed down to 50~K. The absence of detectable superlattice reflections could be associated with the low symmetry of the monoclinic structure and/or peak broadening caused by inherent strain in the crystal. The anomaly at 110~K depicts a crossover between two electronic regimes, wherein the electron-phonon coupling starts freezing out possibly due to the presence of another competing interaction at lower temperatures.

\begin{equation}
\alpha_i = \frac{1}{i_0}\frac{(i-i_0)}{\Delta T}
\label{eq:Thermal_expansion}
\end{equation}

\begin{table}[ht!]
\caption{Comparison of thermal expansion coefficients ($\alpha_i$ for $i$=$a$,$b$, and $c$) in various temperature ranges for monoclinic structure}
\label{tab:structure}
\begin{ruledtabular}
\begin{tabular}{lccc}
Temperature & $\alpha_a$ & $\alpha_b$ & $\alpha_c$ \\
 (K) & (K$^{-1}$) & (K$^{-1}$) & (K$^{-1}$) \\
\hline
310 to 250 & $-9.3(6) \times 10^{-6}$ & $ -5.7(5) \times 10^{-6}$ & $3.5(2) \times 10^{-5}$ \\250 to 110 & $-5(1) \times 10^{-7}$ & $-4(6) \times 10^{-8}$ & $2.14(2) \times 10^{-5}$ \\
110 to 50 & $4.5(7) \times 10^{-6}$ & $1.2(7) \times 10^{-6}$ & $6(2) \times 10^{-6}$ \\
\end{tabular}
\end{ruledtabular}
\end{table}

\subsubsection{Structural properties under external pressure}

\begin{figure*}[htbp]
\centering
\includegraphics[width=\textwidth]{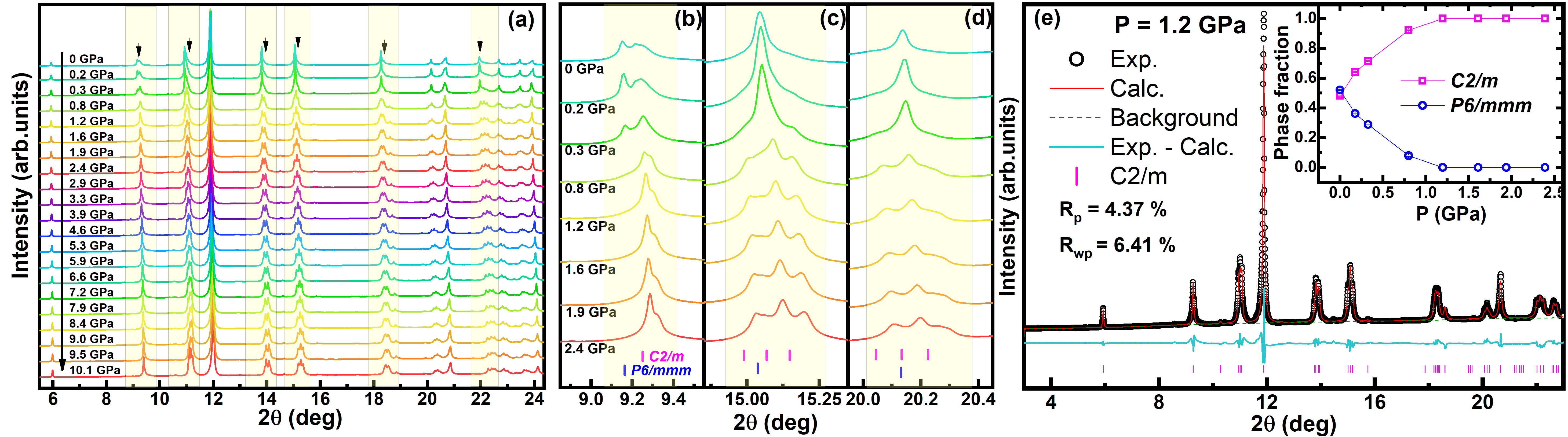}
\caption{(a) Evolution of the XRD pattern under external pressure up to 10~GPa. Structural transition from mixed phase to purely monoclinic phase highlighted around the peak at (b) 9.2$^{\circ}$, (c) 15$^{\circ}$, (d) 20.15$^{\circ}$ at various applied pressures. (e) Synchrotron powder XRD pattern obtained for powdered PIB single crystal at 1.2~GPa at room temperature. The inset in (e) shows the variation of the phase fractions of the two coexisting phases in PIB under application of external pressure.}
\label{fig:PIB_HPrun_PeakEvolution}
\end{figure*}

\begin{figure*}
 \includegraphics[width=\textwidth]{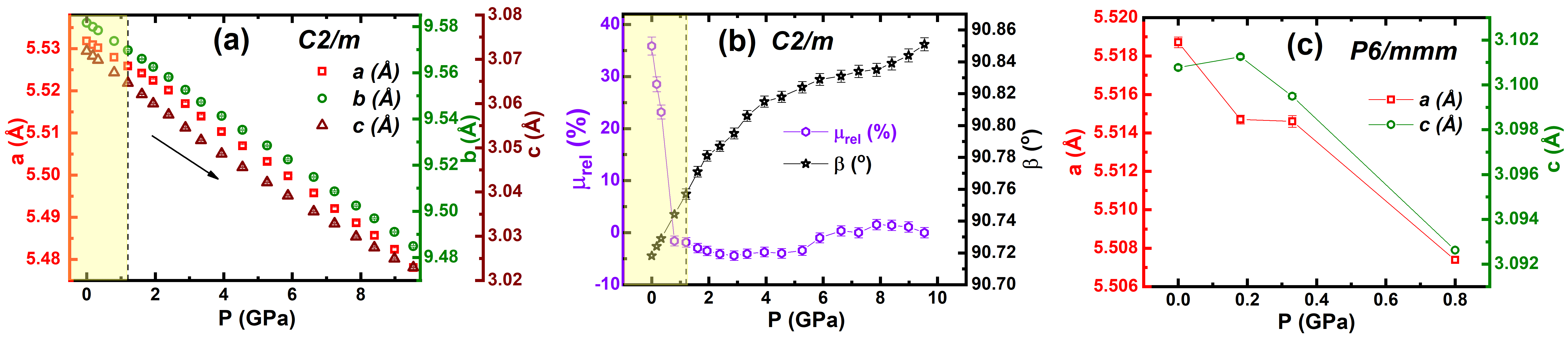}
\caption{Variation of the lattice parameters of monoclinic ((a),(b)) and hexagonal ((c)) phases as a function of pressure obtained from the Rietveld refinement of the pressure-dependent XRD pattern. The shaded region in (a) and (b) indicate the variation of the lattice parameters of monoclinic structure in the mixed phase region.}
\label{fig:S2_Refined_parameters_Pdep}   
\end{figure*}

\begin{figure*}[htb!]
\includegraphics[width=\textwidth]{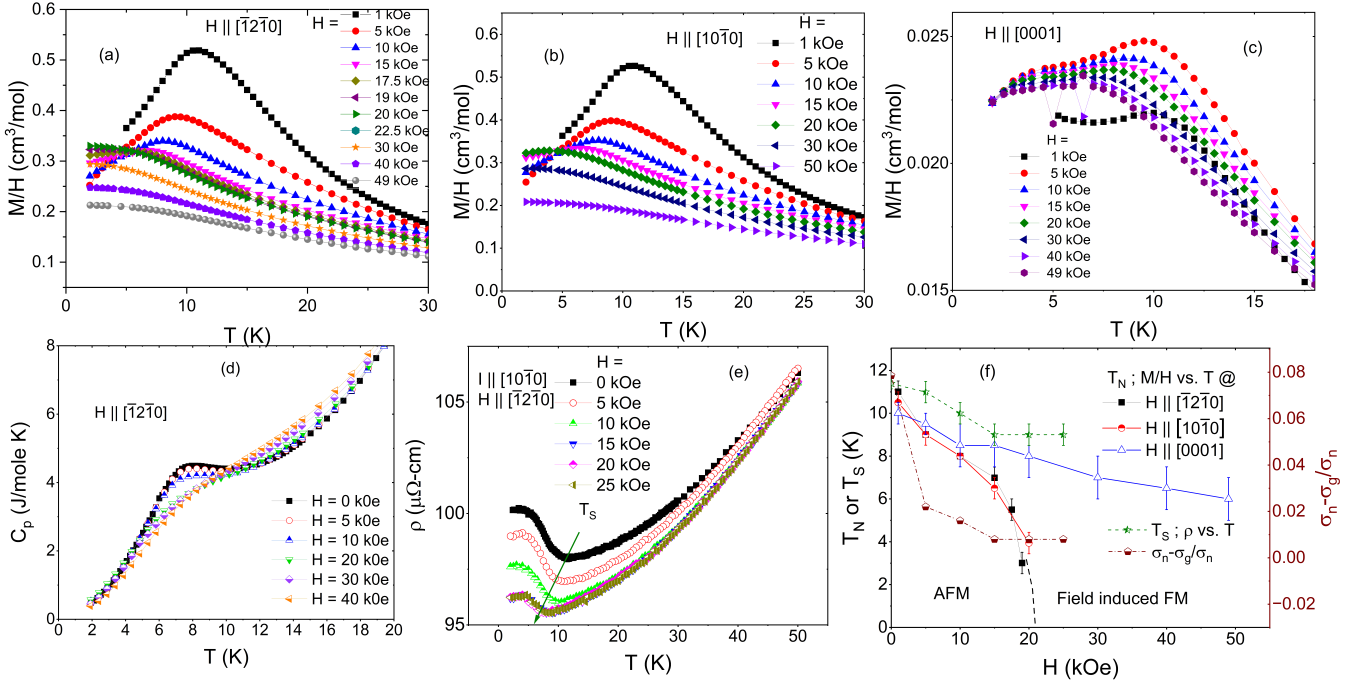}

\caption{\label{mag} $M/H~vs.~T$ at different magnetic field along (a) [$\overline{1}2\overline{1}0$] (b) [$10\overline{1}0$] (c) [0001] directions of \pib.~(d) $C_{\rm P}~vs.~T$ at different magnetic fields along [$\overline{1}2\overline{1}0$] direction. (e)~$\rho~vs.~T$ at different magnetic field parallel to [$\overline{1}2\overline{1}0$] direction, the black arrow marks the shift of superzone gap opening temperature ($T_{\rm S}$). (f) Field dependent phase diagram of \pib: Left axis depicts shift of $T_{\rm N}$ for field along three directions, measured from peak of $M/H~vs.~T$ data and shift of $T_{\rm S}$ with field, measured from the upturn point of $\rho~vs.~T$ data; Right axis depicts change in superzone gap ([$\sigma_n-\sigma_g]/\sigma_n$) with magnetic field along [$\overline{1}2\overline{1}0$] direction.}
\end{figure*}
To understand the effect of external pressure on the crystal structure, pressure-dependent synchrotron XRD measurements were performed at room temperature. Fig.~\ref{fig:PIB_HPrun_PeakEvolution}~(a) shows the evolution of the XRD pattern under external pressure up to 10~GPa. The disappearance of the contribution of the peaks belonging to the hexagonal structure and the appearance of the well-resolved peaks belonging to the monoclinic structure around 1.2~GPa, as highlighted in Figs.~\ref{fig:PIB_HPrun_PeakEvolution}~(b)-(d), clearly signifies the transition into the monoclinic structure. To further confirm the completeness of the transition, Rietveld refinement has been extended to the data collected at higher pressures. 
Under application of external pressure, the phase fraction of the hexagonal phase decreases gradually (shown in the inset of Fig.~\ref{fig:PIB_HPrun_PeakEvolution}~(e)), and the structural transition into the purely monoclinic phase is evident at 1.2~GPa, thus confirming the formation of a single phase monoclinic structure. Therefore, 0~GPa $\leq$ $P <$ 1.2~GPa marks the mixed phase region where the monoclinic and hexagonal phases coexist in a twinned structure and for $P\geq$ 1.2~GPa, the crystal is in the single phase (monoclinic) region. The representative room temperature XRD pattern at 1.2~GPa, refined to a purely monoclinic phase, and at 0.2~GPa, refined to coexisting hexagonal and monoclinic phases, is shown in Fig.~\ref{fig:PIB_HPrun_PeakEvolution}~(e) and Fig.~\ref{fig:P01_refined_rev phase fraction inset} of the supplementary material, respectively. Figs.~\ref{fig:S2_Refined_parameters_Pdep}~(a)-(c) summarises the variation of the structural parameters with external pressure, obtained from the refinement of the XRD patterns for the monoclinic phase from ambient up to 10~GPa and for the hexagonal phase from ambient up to 1.2~GPa. To extract the structural parameters, two-phase refinement using monoclinic and hexagonal phases was employed for $P <$ 1.2~GPa and single-phase refinement using monoclinic structure was used for $P\geq$ 1.2~GPa. An overall decrease in the unit cell volume is observed with the application of pressure, which is expected due to the contraction of the lattice. Unlike the evolution of the lattice parameters with temperature, under external pressure, the $a$, $b$, and $c$ lattice parameters of both the phases vary isotropically (Figs.~\ref{fig:S2_Refined_parameters_Pdep}~(a), (c)). The gradual enhancement in the value of the angle $\beta$ confirms the evolution of the monoclinic structure by the progressive shear distortion of the hexagonal lattice along the $ab$-plane (Fig.~\ref{fig:S2_Refined_parameters_Pdep}~(b)). The variation of $\mu_{\mathrm{rel}}$ (\si{\percent}) (calculated relative to the microstrain at 10~GPa) suggests that the application of pressure releases the inherent strain in the crystal and helps it to attain a more stable monoclinic structure (Fig.~\ref{fig:S2_Refined_parameters_Pdep}~(b)). Although the pressure dependence of the monoclinic phase at ambient temperature is found to have an expected regular compression effects on lattice parameters $a$, $b$, and $c$ and unit-cell volume, a subtle anomaly can be noted above 5~GPa in the monoclinic angle $\beta$ as well as the microstrain. The structural transition was found to be fully reversible under external pressure (see Fig.~\ref{fig:PIB_HPrun_PeakEvolution_rev} and inset of Fig.~\ref{fig:P01_refined_rev phase fraction inset} of the supplementary material). 

Therefore, the application of external pressure is analogous to lowering the temperature in inducing similar structural transition in PIB. This study supports the hypothesis that the electronic transition observed at \SI{\sim 280}{\K} in the single crystal of PrIr$_3$B$_2$ is having a strong correlation to the structural transformation from a mixed phase to a purely monoclinic phase, with a twinned structure present in a broad mixed phase region between 280~K $<T\leq$ 310~K at ambient pressure and between 0~GPa $\leq P <$ 1.2~GPa at room temperature. A structural transition from high-temperature hexagonal to low-temperature monoclinic phase was previously reported in similar systems like CeIr$_3$B$_2$ \cite{Kubota2013}. Nevertheless, to the best of our knowledge, the presence of this broad mixed-phase region containing twinned structure has not been previously identified in this series.

\subsection{Magnetic field tuning of magnetic structure}
Fig.~\ref{mag} (a), (b) and (c) show magnetic susceptibility ($M/H$) $vs.$ temperature ($T$) data, measured at different fields along [$\overline{1}2\overline{1}0$], [$10\overline{1}0$] and [0001] crystallographic directions, respectively, around the antiferromagnetic to paramagnetic phase transitions. For in-plane field along  [$\overline{1}2\overline{1}0$] and [$10\overline{1}0$] directions, the peak in $M/H$ data shifts to lower temperature from 10 to 3~K with increasing magnetic field. This trend is observed up to 20~kOe and beyond that field, no peak is observed in $M/H$ data. A spin-flop transition was indicated above 15~kOe in the earlier single crystal study \cite{Manni2019}. Maybe, the system attains a field-induced canted antiferromagnetic state or a ferromagnetic (FM) state above 20~kOe in-plane field. No observable in-plane magnetic anisotropy is detected. It is also evident from the determined magnetic structure \cite{Sharma2024}. For the magnetic field along the out-of-plane [0001] direction, $M/H$ is one order of magnitude less and shows a weak feature around $T_{\rm N}$, which shifts rather very gradually with increasing field. A transition is observed even at the 50~kOe field at around 6~K. This clearly suggests that the [0001] is the hard direction and moments lie in the $ab$-plane. The change in $T_{\rm N}$ with in-plane magnetic field is investigated by heat capacity ($C_{\rm P}$) $vs.~T$ measurement with $H$ along [$\overline{1}2\overline{1}0$], as shown in Fig.~\ref{mag}~(d). The feature in $C_{\rm P}$ is weak at $T_{\rm N}$, which shifts to lower temperature upto 10~kOe field. From a 20~kOe field and beyond, the peak broadens, and entropy is distributed to higher temperatures. This confirms the field-induced polarized state above 20~kOe in-plane field.  

Fig.~\ref{mag}~(e) shows variation of superzone gap size and the gap opening temperature ($T_{\rm S}$) with increasing field along [$\overline{1}2\overline{1}0$] direction from the temperature-dependent resistivity ($\rho$) measurement. The current is applied along [$10\overline{1}0$] direction. $T_{\rm S}$ is the temperature at which resistivity starts increasing upon lowering the temperature. At zero field, it is equal to 11.4~K, very close to $T_{\rm N}$. Upon increasing the field to 15~kOe, it shifts to 9~K and beyond that, no shift of $T_{\rm S}$ is observed. The superzone gap is calculated by ($\sigma_n - \sigma_g$)/$\sigma_n$, where $\sigma_n$, and $\sigma_g$ are normal state and gapped conductivities, reciprocal of the respective resistivities: $\rho_n$ and $\rho_g$. The $\rho_n$ is the resistivity at $T_{\rm S}$ and $\rho_g$ is the maximum resistivity below $T_{\rm S}$. It has been observed that for in-plane magnetic field upto 15~kOe, the size of the superzone gap decreases from 0.075 to 0.008 (one order of magnitude) when the $T_{\rm N}$ is also suppressed by more than three times. It is noteworthy that the superzone gap at the Fermi surface persists in the magnetically polarized state above 20~kOe, unchanged at a value of 0.008. 

Fig.~\ref{mag}~(f) summarises the tunability of the magnetic structure of \pib. Sharma et. al. proposed two collinear magnetic structure for \pib: (i) along [100] direction of $C2/m$ crystal structure or [$10\overline{1}0$] direction of $P6/mmm$ structure, (ii) at a 42$^{\circ}$ angle from [$10\overline{1}0$] direction; in the $ab$-plane \cite{Sharma2024}. The absence of any in-plane anisotropy in the field-dependent variation of $T_{\rm N}$ and magnitude of $M/H$ and nature of variation of the same for the out-of-plane field clearly suggests that the moments are antiferromagnetically arranged in the $ab$-plane and they point at an angle with principle crystallographic axis, may be at a 45$^{\circ}$ angle. This finding agrees with (1/2, 1/2, 0) magnetic propagation vector suggested by Sharma et. al \cite{Sharma2024}. The moments attain a polarized ferromagnetic state above 20~kOe in-plane field.  It is interesting to see that the size of the superzone gap strongly suppresses with decrease of $T_{\rm N}$ with the in-plane field. This suggests the mechanism of superzone gap opening in the Fermi surface is strongly coupled with magnetic order \cite{ZHANG2004, Das_2012, Green2025}. The fact that the superzone gap opening doesn't close even in the fully polarized state indicates that the superzone gap acts on the remnant, ungapped metallic pockets of the FS left behind by the CDW phase transition \cite{Ru2008, Regmi2023}. Hence even a small residual exchange field is sufficient to keep those pockets gapped out. 

\section{Conclusion}
The single crystal of PrIr$_3$B$_2$ grown by the Czochralski pulling method is found to exist in a twinned structure of monoclinic and hexagonal phases at ambient conditions of temperature and pressure. The electrical resistivity measurement performed on the single crystal revealed a two-step metal-insulator-metal transition (at \SI{\sim 280} {\K} and \SI{\sim 250} {\K}) along with a linear temperature dependent resistivity behaviour between 110 K $<T\leq$ 250~K, a deviation from linear behavior below \SI{\sim 110} {\K}, and a resistive upturn below \SI{\sim 10} {\K}. The transition at 280~K is found to be structural in origin from the high-temperature hexagonal ($P6/mmm$) to the low-temperature monoclinic ($C2/m$) structure. This change in the crystal symmetry is found to be gradual and progressive with a broad twinned structure regime likely due to inherent strain in the pulled crystal. The transition is pressure-tunable. As the temperature (pressure) is decreased (increased), the monoclinic structure is found to evolve at the expense of the hexagonal structure by the shear distortion of the lattice in the $ab$-plane. The second resistive transition observed at \SI{\sim 250}{\K} in the cooling cycle, and the deviation from the linear resistivity below 110~K in the transport measurement have noticeable changes in the structural parameters of the monoclinic structure without appearance of any additional peak, suggesting the transitions are electronic in origin mediated by strong electron-phonon coupling. We have proposed that Fermi surface is partially gapped below 250~K due to possible charge ordered state which leads to linear resistivity. The superzone gap due to AFM ordering at \SI{\sim 10} {\K} does not close in the polarized state and may be connected to the gapped region of the Fermi surface opened at around 250~K. 

In Summary, this study unfolds multiple structural and electronic transitions in the kagome lattice compound PrIr$_3$B$_2$ and provides a very plausible explanation of the unusual two-step MIM transition, hinting towards a CDW state. Moreover, the magnetic phase diagram is also mapped, which shows a possible connection between the low-temperature magnetic transition and the high-temperature MIM transition.

\section{Data availability}
Data sets generated during the current study are available from the corresponding
author on reasonable request.

\section{Competing interests}
The authors declare no competing interests.

\section{Acknowledgement}
Authors acknowledge CIF, IIT Palakkad and SAIF, IIT Madras for experimental facilities. Authors express sincere gratitude to Prof. Arumugam Thamizhavel and Prof. Sudesh K. Dhar for fruitful discussion, experimental support for the magnetic phase diagram study, and suggestions on the manuscript, and Ruta Kulkarni for help in single crystal growth. G.V. and S.M. thank the Department of Science \& Technology (DST), Government of India (GoI) for funding of the user travel and expenses for the Elettra beamline proposal 20245241, and the Xpress beamline, Elettra Sincrotrone Trieste, Italy, for providing all user support. G.V. acknowledges the DST, GoI, for financial support vide reference no. DST/WISE-PDF/PM-82/2023 under the WISE Post-Doctoral Fellowship programme to carry out this work. K.A.I. acknowledge the IISc, Bengaluru and ICTP, Trieste for the IISc/ICTP fellowship.

\section{Author contributions}
S.M. has grown the single crystal. G.V., K.A.I., and B.J. carried out the synchrotron XRD experiments at Elettra, and S.K.M. carried out the initial synchrotron experiments in another beamline (data not used here). G.V., K.A.I. and S.K.M. analyzed the XRD data. DSC data was analyzed by G.V.  S.M. carried out the experiments for the magnetic phase diagram and analyzed the data. G.V., K.A.I. and S.M. wrote the main manuscript text. G.V., S.K.M.  and S.M. (Figure 7) prepared the figures. All authors reviewed the manuscript.

\section{References}

\bibliographystyle{apsrev4-2}

\newpage
\section{Supplementary Material}
\subsection{Differential Scanning Calorimetry (DSC)}

\begin{figure}[htbp]
\centering
\includegraphics[width=\columnwidth]{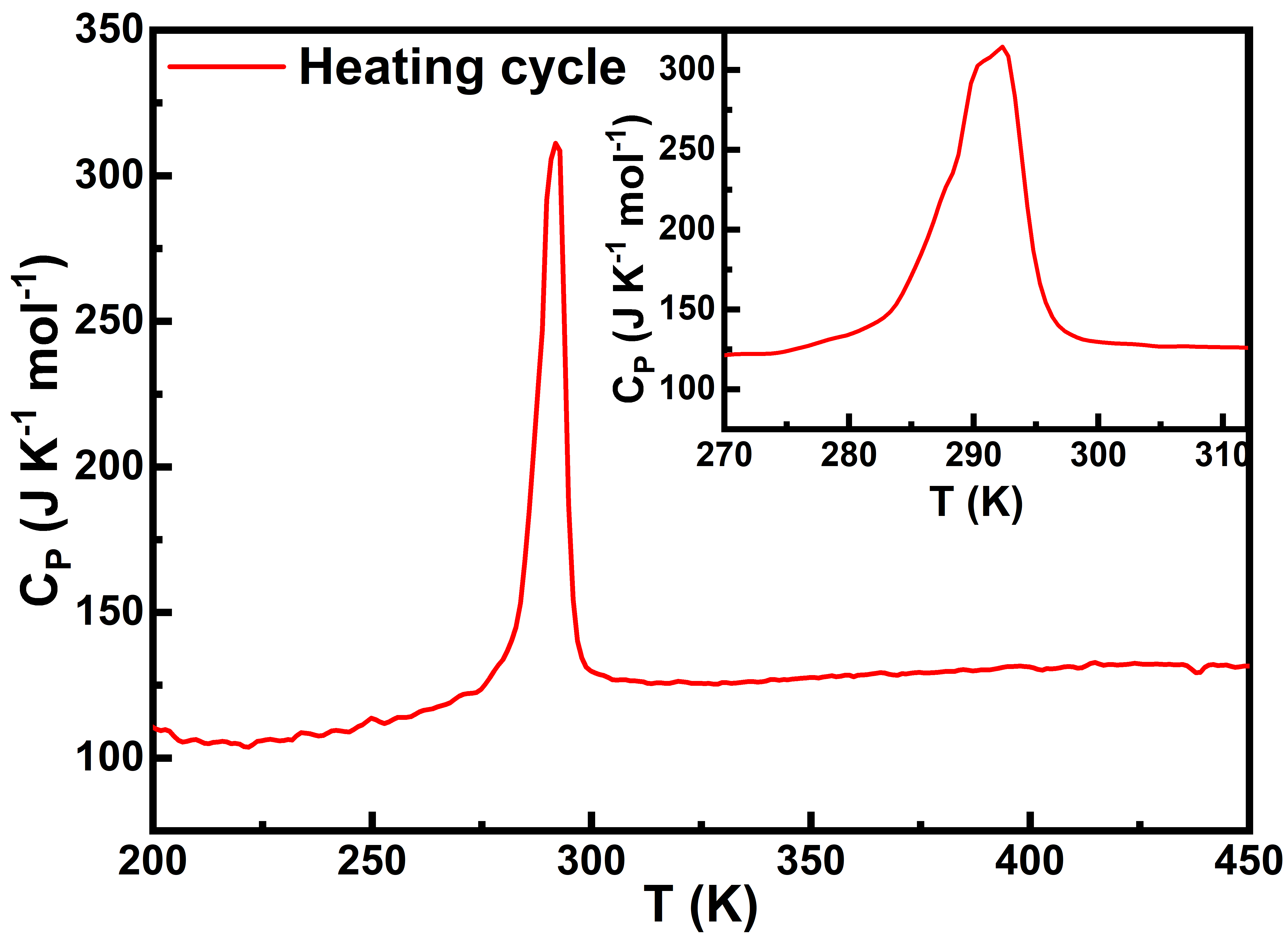}
\caption{DSC specific heat capacity in the heating cycle}
\label{fig:DSC_2a}
\end{figure}

Fig.~\ref{fig:DSC_2a} shows the DSC heat capacity curve obtained for PrIr$_3$B$_2$ (PIB) single crystal. A broad peak between 285~K and 295~K with a jump of around 200~J/K-mol is highlighted in the inset.

\subsection{Electrical resistivity}

\begin{figure}[htbp]
\centering
\includegraphics[width=\columnwidth]{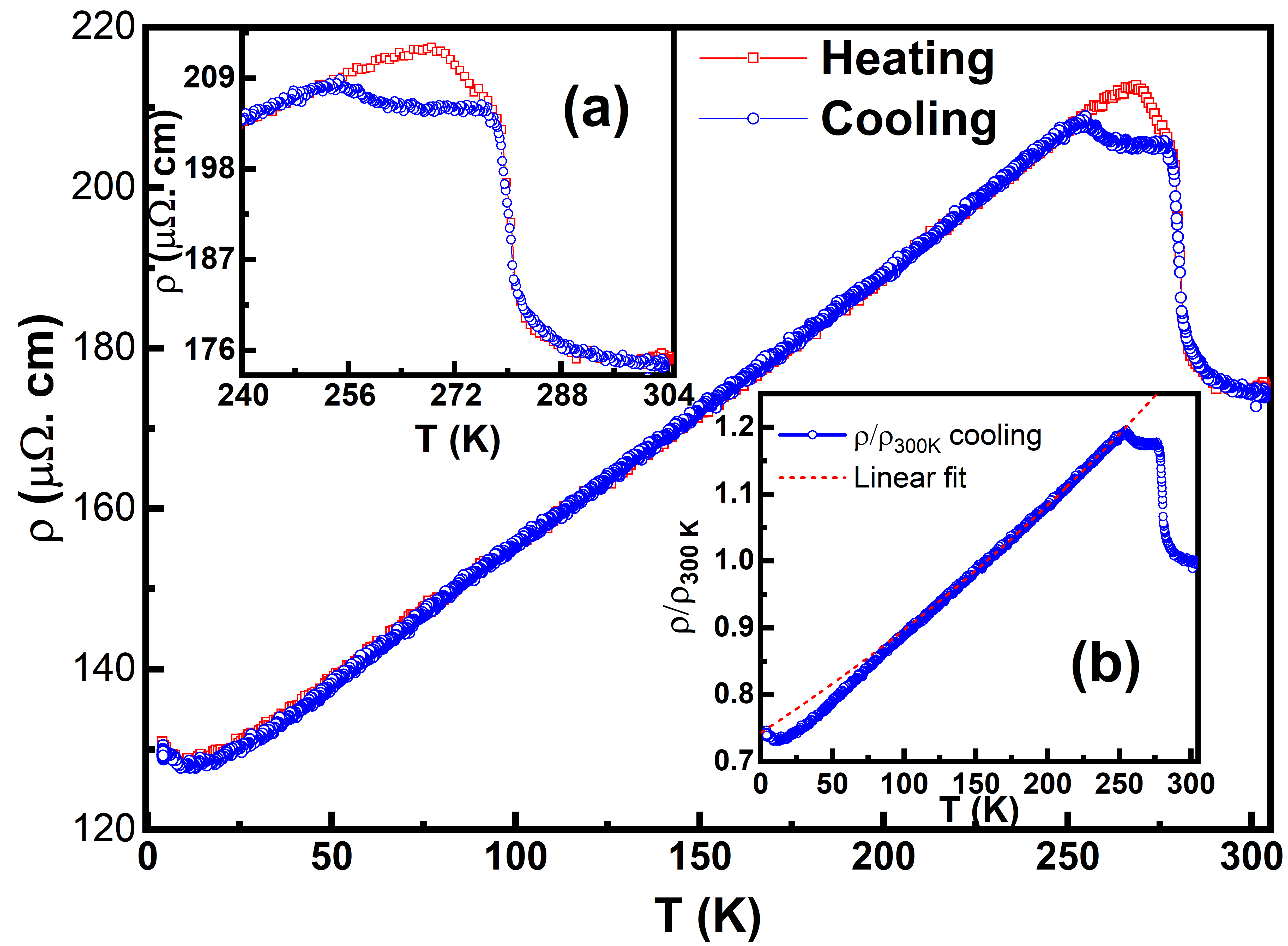}
\caption{Temperature-dependent electrical resistivity ($\rho(T)$ vs $T$) in the heating and cooling cycles. Inset (a) highlights the two-step metal-insulator-metal transition in the temperature range 250~K to 280~K. Inset (b) indicates the linear fit of the plot $\rho/\rho_{300~K}$ vs $T$.}
\label{fig:RT_insets_ab_linear fit}
\end{figure}

Fig.~\ref{fig:RT_insets_ab_linear fit} shows the electrical resistivity curve obtained for PIB single crystal. An anomalous two-step metal-insulator-metal transition in the temperature range 250~K to 280~K is highlighted in the inset (a). Inset (b) portray the linear fit of $\rho/\rho_{300~K}$ vs $T$ in the T-range 250~K $\geq T>$ 110~K and an anomaly at \SI{\sim 110}{\K} in the form of deviation from this linear behaviour.

\subsection{Temperature and pressure tuning of the crystal structure}

The crystallographic data (structural parameters and atomic coordinates) for the two coexisting phases in PrIr$_3$B$_2$ obtained from Reitveld refinement analyses at ambient pressure are provided in Table~\ref{tab:PrIr3B2}.

\begin{table*}[t]
\caption{Structural parameters and atomic coordinates of PrIr$_3$B$_2$
obtained from Rietveld refinement at ambient pressure.}
\label{tab:PrIr3B2}
\centering

%--------------------------------------------------
% Structural parameters
%--------------------------------------------------

\begin{tabular}{lcc}
\hline
\multicolumn{3}{l}{\textbf{Chemical formula:} PrIr$_3$B$_2$} \\

\multicolumn{3}{l}{\textbf{Molecular weight:} 739.18 g/mol} \\

\multicolumn{3}{l}{\textbf{Radiation type:} Synchrotron,
$\lambda = 0.4953$ \AA} \\

\multicolumn{3}{l}{\textbf{Pressure:} $\sim$0 GPa} \\

\hline

\textbf{Temperature} & \textbf{300 K} & \textbf{280 K} \\

\textbf{Crystal system} & \textbf{Hexagonal} & \textbf{Monoclinic} \\

\textbf{Space group} & $P6/mmm$ & $C2/m$ \\

\textbf{$a$ (\AA)} & 5.5179(3) & 5.53397(8) \\

\textbf{$b$ (\AA)} & 5.5179(5) & 9.5822(1) \\

\textbf{$c$ (\AA)} & 3.0992(1) & 3.0703(1) \\

\textbf{$V$ (\AA$^3$)} & 81.719(4) & 162.801(6) \\

\textbf{$\beta$ ($^\circ$)} & 90 & 90.689(4) \\

\hline
\end{tabular}

\vspace{3mm}

%--------------------------------------------------
% Atomic coordinates
%--------------------------------------------------

\begin{tabular}{llccc llccc}
\hline

\multicolumn{10}{c}{\textbf{Atomic coordinates}} \\[2mm]

&
&
\multicolumn{3}{c}{\textbf{Hexagonal}}
&
&
&
\multicolumn{3}{c}{\textbf{Monoclinic}} \\

\textbf{Atom}
&
\textbf{Wyckoff}
&
\textbf{x}
&
\textbf{y}
&
\textbf{z}
&
\textbf{Atom}
&
\textbf{Wyckoff}
&
\textbf{x}
&
\textbf{y}
&
\textbf{z}
\\

&
\textbf{Position}
&
&
&
&
&
\textbf{Position}
&
&
&
\\

\textbf{Ir}
& 3g
& 1/2
& 0
& 1/2
&
\textbf{B}
& 4h
& 0
& 0.23307
& 1/2
\\

\textbf{B}
& 2c
& 1/3
& 2/3
& 0
&
\textbf{Ir1}
& 4e
& 1/4
& 0
& 1/4
\\

\textbf{Pr}
& 1a
& 0
& 0
& 0
&
\textbf{Pr}
& 2d
& 0
& 1/2
& 1/2
\\

$-$
& $-$
& $-$
& $-$
& $-$
&
\textbf{Ir2}
& 2a
& 0
& 0
& 0
\\

\hline
\end{tabular}

\end{table*}

\begin{figure}[htbp]
\centering
\includegraphics[width=\columnwidth]{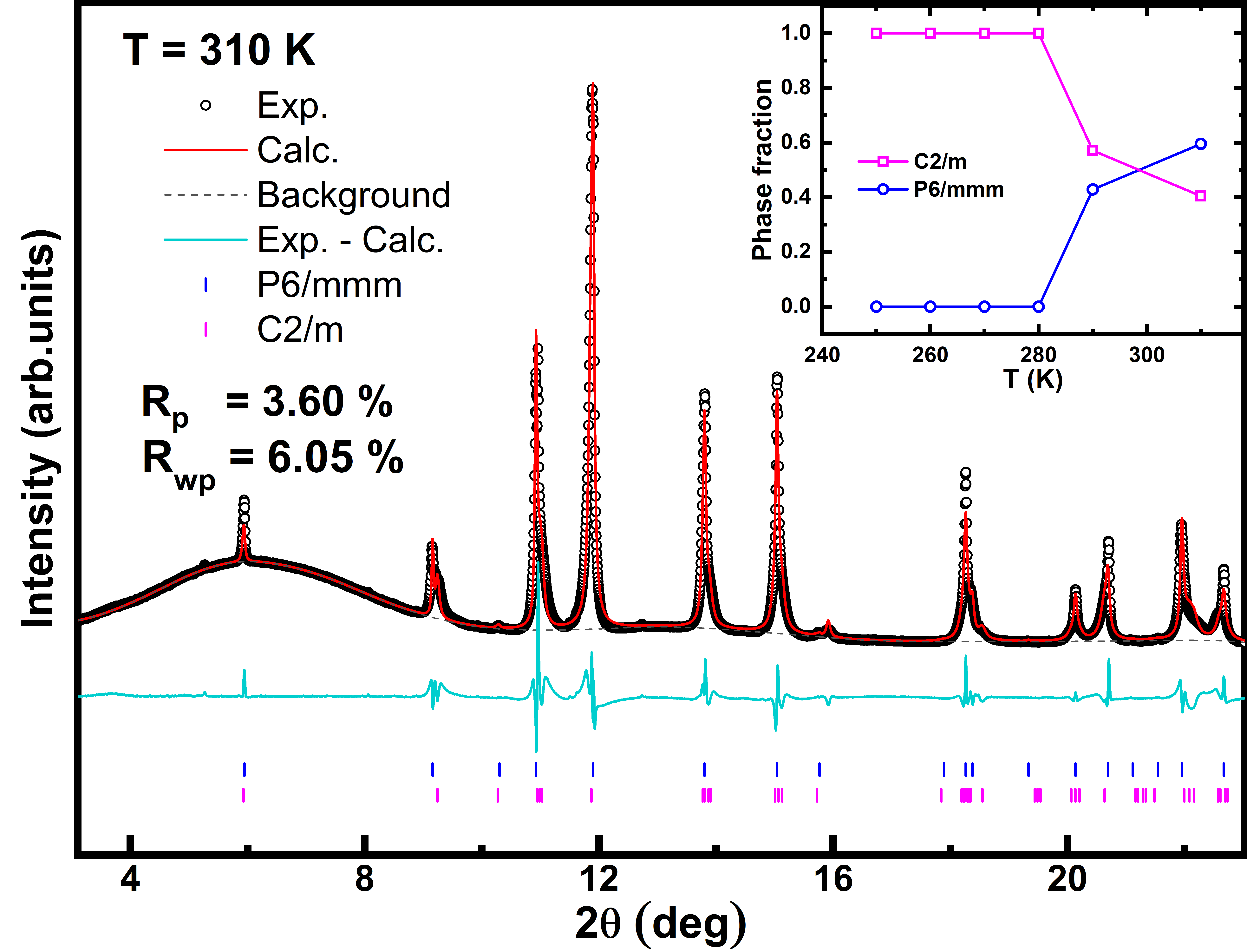}
\caption{Synchrotron powder XRD pattern obtained for powdered single crystals at 310 K and ambient pressure. The inset shows the variation of phase fraction with temperature in the heating cycle.
}
\label{fig:310K_refinement_PF_Heating}
\end{figure}

\begin{figure}[htbp]
\centering
\includegraphics[width=\columnwidth]{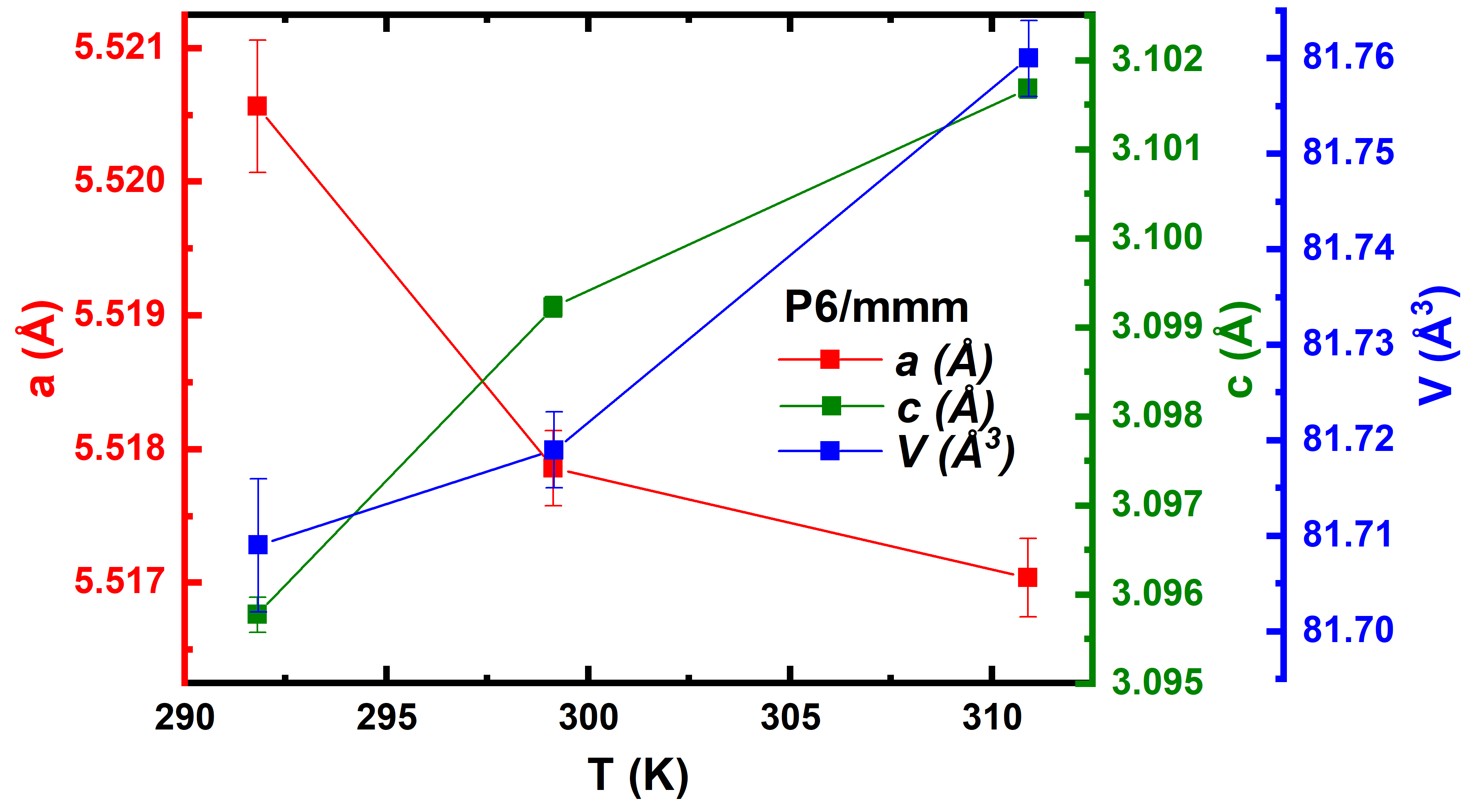}
\caption{Variation of the lattice parameters of the hexagonal phase as a function of temperature obtained from the Rietveld refinement of the temperature-dependent XRD pattern.
}
\label{fig:lattice_parameter_LT_Hex}
\end{figure}

\begin{figure}[htbp]
\centering
\includegraphics[width=\columnwidth]{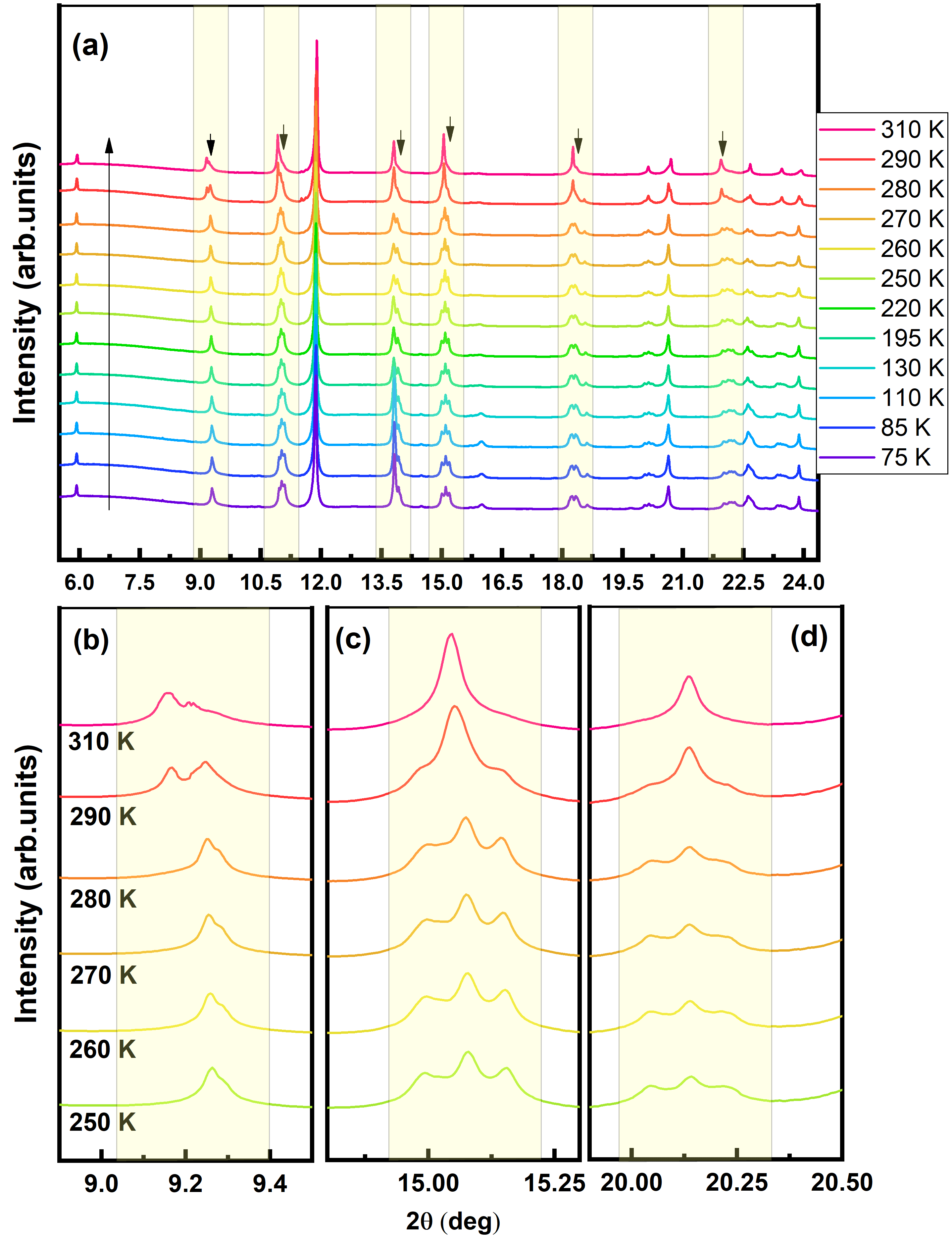}
\caption{(a) Evolution of the XRD pattern under heating cycle. Structural transition from purely monoclinic phase to the mixed phase is highlighted around the peak at (b) 9.2$^{\circ}$, (c) 15$^{\circ}$, (d) 20.15$^{\circ}$ at various pressures. The $(331)$, $(-331)$ and $(061)$ peaks of the $C2/m$ structure gradually merges into the $(301)$ peak of $P6/mmm$ structure in (d).}
\label{fig:PIB_LTrun_PeakEvolution_FWHM_1}
\end{figure}

Fig.~\ref{fig:310K_refinement_PF_Heating} represents the synchrotron powder XRD pattern obtained for the powdered single crystal of PIB at 310~K and ambient pressure. The variation of the phase fraction of the coexisting phases in the heating cycle is illustrated in the inset.

Fig.~\ref{fig:lattice_parameter_LT_Hex} illustrates the variation of the lattice parameters of the hexagonal phases as a function of temperature obtained from the Rietveld refinement of the temperature-dependent XRD pattern. The lattice parameters of the monoclinic structure in the mixed phase region were obtained by extrapolating its trend from the single-phase region ($T \leq$ 280~K) and were kept fixed to obtain the refined hexagonal lattice parameters.

Fig.~\ref{fig:PIB_LTrun_PeakEvolution_FWHM_1}~(a) shows the evolution of the XRD pattern under heating cycle illustrating the reversibility of the transition. In the reverse (heating) cycle, the $(331)$, $(-331)$, and $(061)$ peaks of the $C2/m$ structure gradually merges into the $(301)$ peak of $P6/mmm$ structure (Fig.~\ref{fig:PIB_LTrun_PeakEvolution_FWHM_1}~(c)-(d)). Above 280~K, the peaks of the $P6/mmm$ structure start emerging (Fig.~\ref{fig:PIB_LTrun_PeakEvolution_FWHM_1}~(b)).

\begin{figure}[htbp]
\centering
\includegraphics[width=\columnwidth]{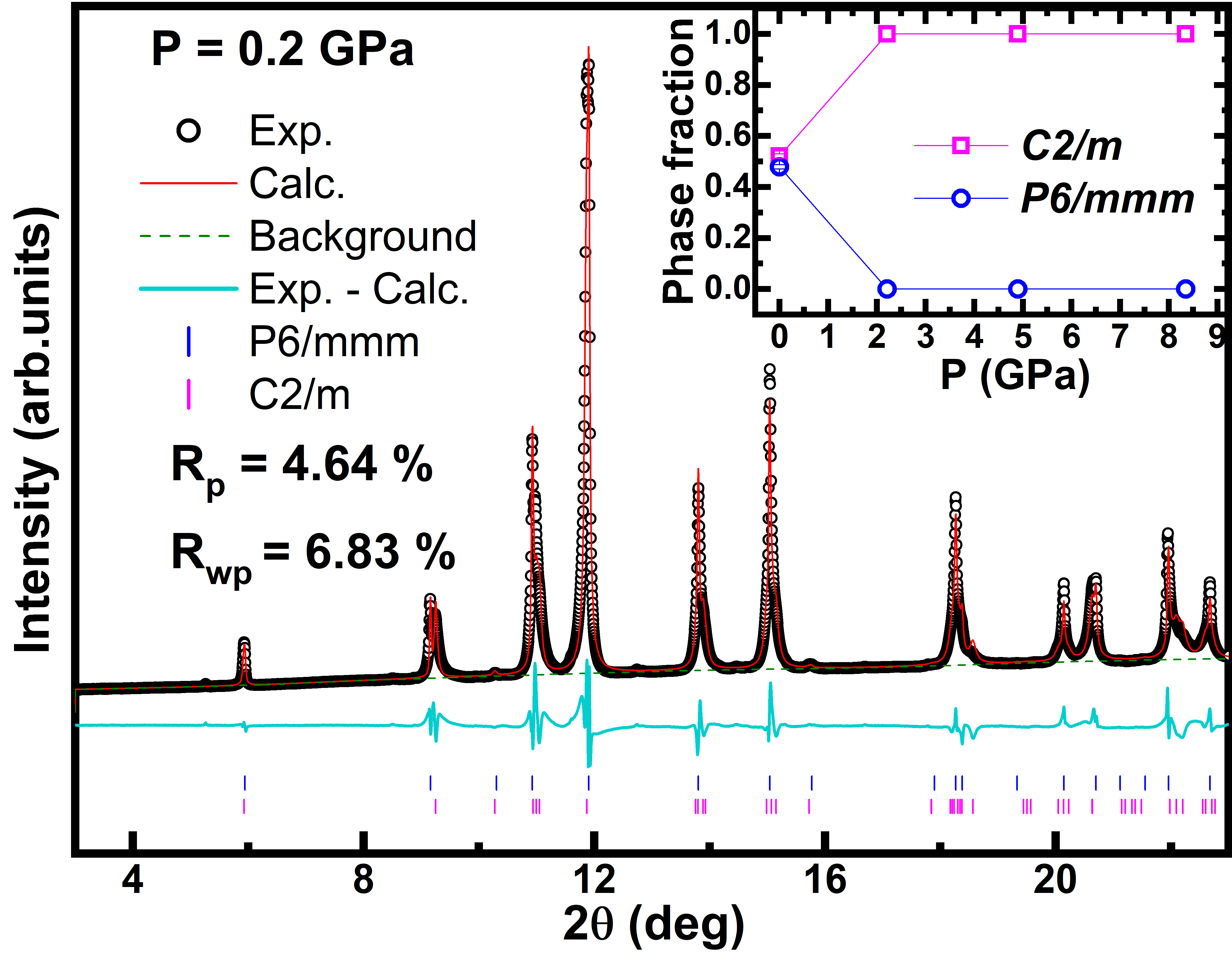}
\caption{Synchrotron powder XRD pattern obtained for pow-
dered PIB single crystal at 0.2 GPa at
room temperature. Inset shows the variation of the phase fractions of the two coexisting phases in PIB in the pressure-releasing (reverse) cycle.}
\label{fig:P01_refined_rev phase fraction inset}
\end{figure}

\begin{figure}[htbp]
\centering
\includegraphics[width=\columnwidth]{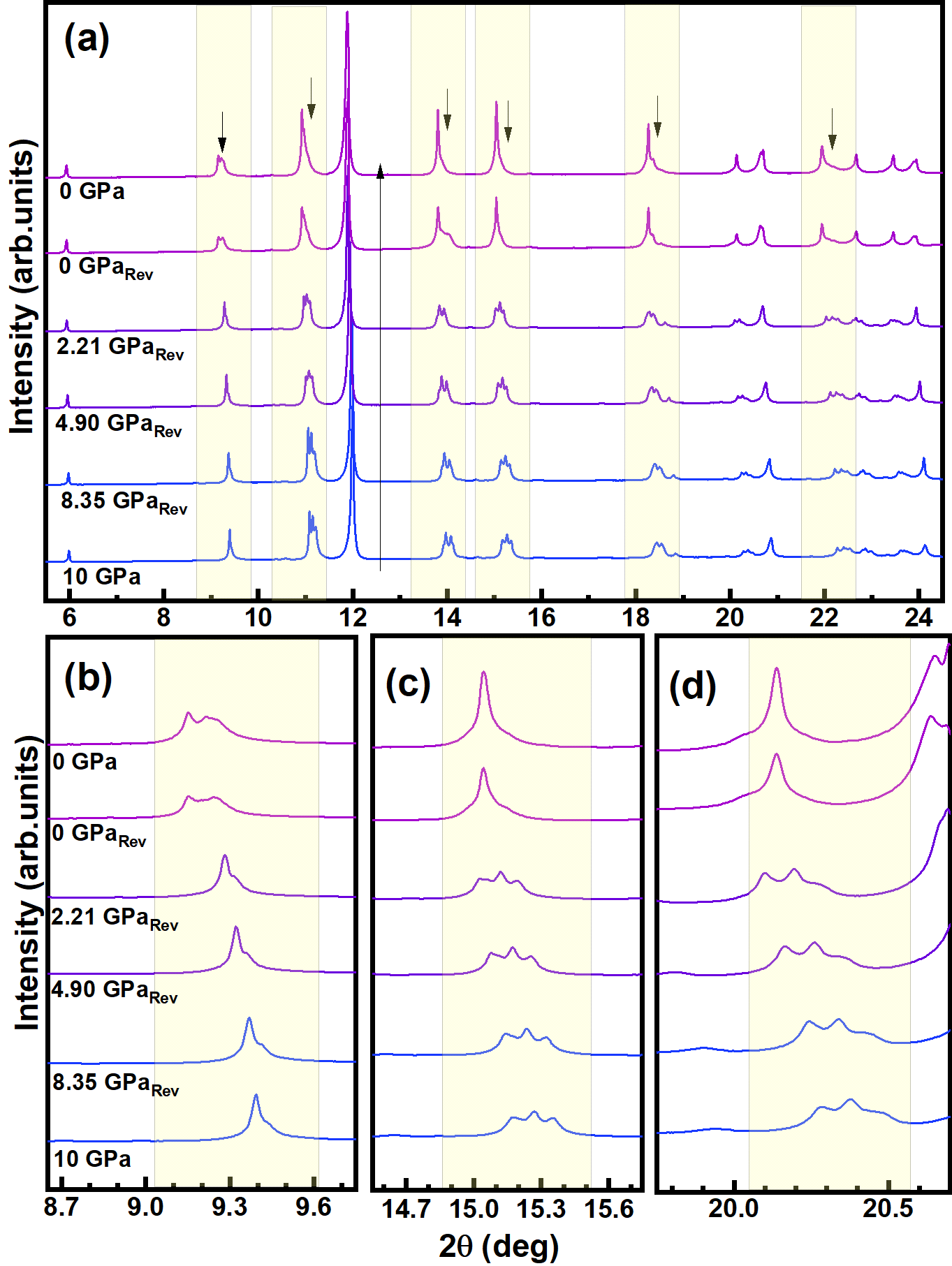}
\caption{(a) Evolution of the XRD pattern under external pressure in the reverse cycle. Structural transition from purely monoclinic phase to the mixed phase is highlighted around the peak at (b) 9.2$^{\circ}$, (c) 15$^{\circ}$, (d) 20.15$^{\circ}$ at various pressures. The $(331)$, $(-331)$ and $(061)$ peaks of the $C2/m$ structure gradually merges into the $(301)$ peak of $P6/mmm$ structure in (d).}
\label{fig:PIB_HPrun_PeakEvolution_rev}
\end{figure}

Fig.~\ref{fig:P01_refined_rev phase fraction inset} represents the room temperature synchrotron powder XRD pattern obtained for powdered single crystal of PIB at 0.2~GPa, refined to coexisting hexagonal and
monoclinic phases. The variation of the phase fraction of the coexisting phases in the reverse cycle is illustrated in the inset.

Fig.~\ref{fig:PIB_HPrun_PeakEvolution_rev}~(a) shows the evolution of the XRD pattern under pressure reverse cycle illustrating the reversibility of the transition. The subscript rev is used to indicate the data collected in the reverse cycle. In the reverse cycle, the $(331)$, $(-331)$ and $(061)$ peaks of the $C2/m$ structure gradually merges into the $(301)$ peak of $P6/mmm$ structure (Fig.~\ref{fig:PIB_HPrun_PeakEvolution_rev}~(c)-(d)). Below 1.2~GPa, the peaks of the $P6/mmm$ structure start emerging (Fig.~\ref{fig:PIB_HPrun_PeakEvolution_rev}~(b)). The ambient pattern in both the cycles are in good agreement.

\begin{figure}[htbp]
\centering
\includegraphics[width=\columnwidth]{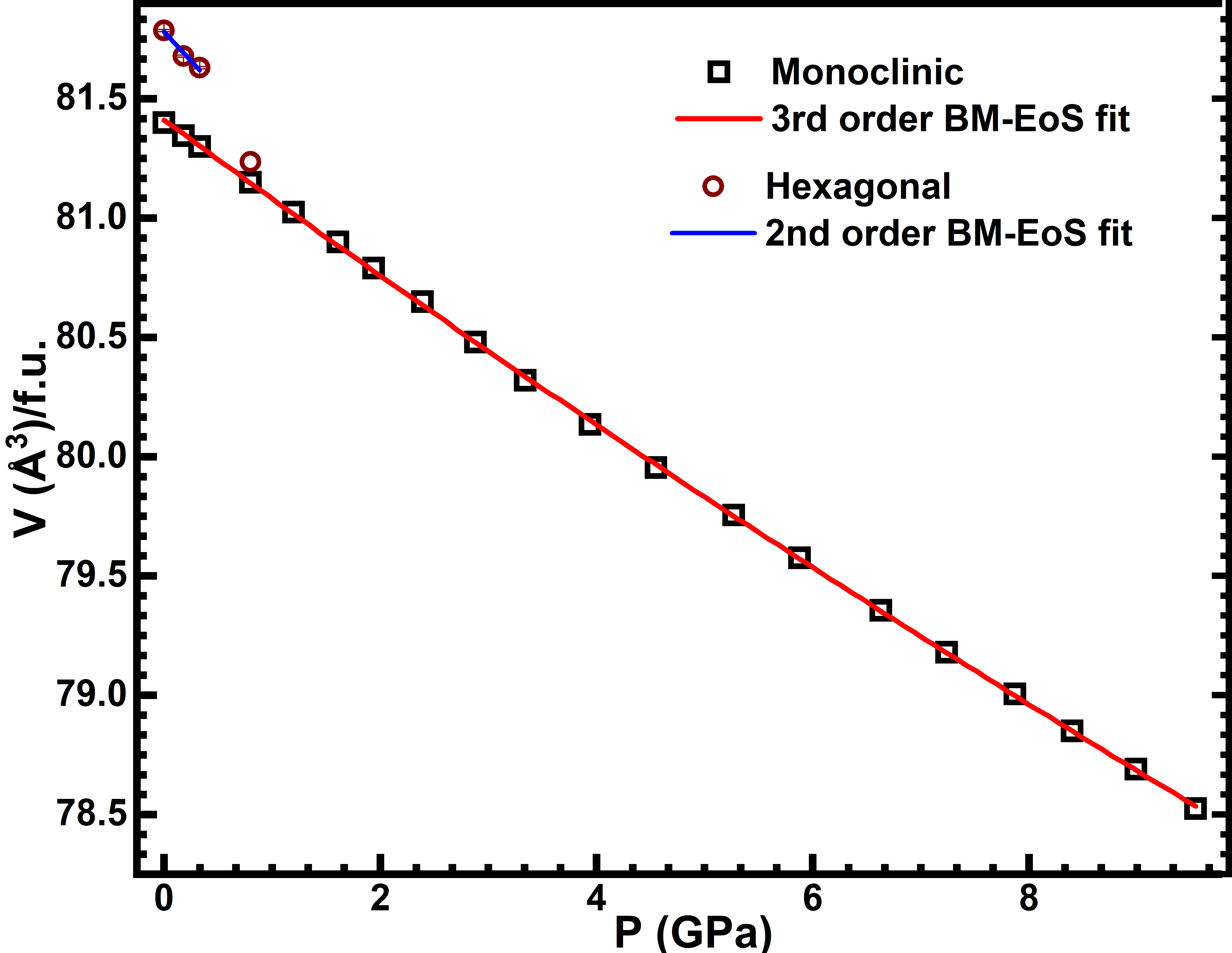}
\caption{P-V curve along with the BM-EoS fit obtained for the monoclinic and the hexagonal phases}
\label{fig:BM_EoS_fit}
\end{figure}

The pressure-volume ($P-V$) curve obtained for the monoclinic and the hexagonal phases in PIB from the refinement of the XRD pattern was further analyzed using Birch-Murnaghan equation of state (BM-EoS). Fig.~\ref{fig:BM_EoS_fit} represent the P-V curves obtained for the monoclinic and the hexagonal phases in PIB, analyzed within the realm of the third-order and the second-order BM-EoS, respectively. The solid red line represents the BM-EoS fit to the $P-V$ curve. The second-order and third-order BM-EoS are given by equations \eqref{eq:BM2_EoS} and \eqref{eq:BM3_EoS}, respectively.
\begin{equation}
P(V) = \frac{3}{2} B_0 
\left[
\left( \frac{V_0}{V} \right)^{7/3}
-
\left( \frac{V_0}{V} \right)^{5/3}
\right]
\label{eq:BM2_EoS}
\end{equation}

\begin{equation}
\begin{aligned}
P(V) &= \frac{3}{2} B_0 
\left[
\left( \frac{V_0}{V} \right)^{7/3}
-
\left( \frac{V_0}{V} \right)^{5/3}
\right] \\
&\quad \times
\left\{
1 + \frac{3}{4}(B_0' - 4)
\left[
\left( \frac{V_0}{V} \right)^{2/3} - 1
\right]
\right\}
\end{aligned}
\label{eq:BM3_EoS}
\end{equation}
Here, $B_0$ is the isothermal bulk modulus, $B_0'$ is its pressure derivative at ambient temperature and pressure, and $V_0$ and V are the volumes at ambient pressure and at a pressure, respectively. The values of the bulk moduli, as calculated for the monoclinic and the hexagonal phases in PIB were found to be $B_{0M}$=243(2)~GPa with $B_{0M}'$ = 4.7(6)~and $B_{0H}$=166(26)~GPa, respectively. The larger value of the bulk modulus obtained for the monoclinic phase indicates its lower compressibility and stronger bonding compared to that of the hexagonal phase. These values of the experimentally obtained bulk moduli are in reasonably good agreement with those reported for other iso-structural systems \cite{D5MA00091B}. 

\end{document}